# Nose Heat: Exploring Stress-induced Nasal Thermal Variability through Mobile Thermal Imaging

Youngjun Cho[1*], Nadia Bianchi-Berthouze[2], Manuel Oliveira[3], Catherine Holloway[1], Simon Julier[1]
[1]Department of Computer Science, University College London (UCL), United Kingdom
[2]Division of Psychology and Language Sciences, University College London (UCL), United Kingdom
[3]Sintef, Norway
[*]corresponding author: youngjun.cho@ucl.ac.uk

*Abstract*—Automatically monitoring and quantifying stress-induced thermal dynamic information in real-world settings is an extremely important but challenging problem. In this paper, we explore whether we can use mobile thermal imaging to measure the rich physiological cues of mental stress that can be deduced from a person's nose temperature. To answer this question we build i) a framework for monitoring nasal thermal variable patterns continuously and ii) a novel set of thermal variability metrics to capture a richness of the dynamic information. We evaluated our approach in a series of studies including laboratory-based psychosocial stress-induction tasks and real-world factory settings. We demonstrate our approach has the potential for assessing stress responses beyond controlled laboratory settings.

## I. Introduction

As humans are homeothermic, our internal temperature is strongly linked to numerous physiological and psychological mechanisms. Given this, human thermal patterns have been widely explored as a way to improve the understanding of our bodies and minds. Amongst temperature monitoring channels, thermal imaging has been shown to be highly effective. In medical applications, for example, it has been used to detect pathological symptoms and disorders in a contactless manner [1]–[3]. Recent advances in commercial thermal imaging technologies have made it possible to use this approach in human computer interaction (HCI) beyond the highly constrained situation of a medical environment [4], [5]. In particular, it has been used to investigate physiological thermal signatures for assessing a person's psychological states, in particular, affective states [6]–[12].

Among the different physiological activities, vasoconstriction and vasodilation patterns underneath a person's skin can be captured through thermal imaging. Such patterns cause increases or decreases in blood flow, contributing to skin temperature changes. Such physiological activities are influenced not only by ambient temperatures (e.g. local cooling or warming) [13], but also by mental stress [14]. Hence, studies have attempted to capture stress or mental workload-induced thermal directional changes, particularly from facial regions of interest (ROIs), while controlling environmental temperature [7]–[9], [15], [16]. Amongst other facial areas, the nose tip has been shown to be the only consistently reported region where we can monitor significant decreases in its temperature under stress conditions, indicating that the nasal thermal drop could be a stress indicator.

Despite this promising finding and a range of low-cost, small thermal cameras, which are already available in the market, these observations have not caught much attention in real-world applications. This is mainly because the majority of thermal imaging-based studies have drawn upon visual inspections (e.g. manually selecting a pair of thermal images where a person's nose is situated in the same position to compare nose temperature) and also imposed motion constraints (e.g. using a chinrest), which is highly cumbersome [8], [9], [11]. To address these restrictions, [16] used a traditional ROI tracking algorithm; however, participants were still required to keep their head. This inflexibility is one of the main challenges for real-world applications and contributes to keeping thermal imaging from being used in unconstrained, mobile settings.

The use of advanced ROI tracking methods built for mobile thermal imaging in unconstrained situations, such as Thermal Gradient Flow [5], can help to address this challenge as it was preliminary explored for very short period of time in [6]. However, the nose tip area is an area which is difficult to track on thermal images. For instance, Fig. 1 shows examples of thermal images of a person's face collected from an office and the nose tip area as a ROI. The shape of the local facial skin region is often blurred. This does not provide a sufficient number of key facial features which are generally required for ROI tracking [5]. This is due to the homoeothermic metabolism and relatively low thermal conductivity of a person's skin (here, the nose tip) which results in a very narrow range of temperature distributions across the nose tip. Also, the thermal patterns on the nose tip are not consistent during the course of time - they are highly variable in comparison with other facial temperatures [9]. This leads to our first research question: *how can we continuously monitor thermal variable patterns of the nose tip in unconstrained settings?*

A second challenge is how to enrich the quality of information derived from affect-induced skin temperature changes. In the literature, *temperature difference* or *slope* between two time points on a facial ROI (positive or negative direction and its amplitude in temperature change, e.g. average of -0.56°C from the nasal ROI after being exposed to stressors in [7]) has been used as the dominant metric. However, a person's skin temperature could be influenced by many factors. These include: different physiological phenomenon (e.g. nose temperature can be affected by not only vasoconstriction/dilation but also breathing [6]), other types of affective states (e.g. nasal temperature drop under fear conditions [10]), context (e.g. social contact increases nasal temperature [17]) and environmental temperature [6]. All of these have the potential to induce temperature variability rather than just a consistent drop in temperature (ie. a simple directionality change). Hence, using temperature difference as a single metric in unconstrained setting could be too sensitive to other factors beyond stress-induced physiological reactions. Indeed, incongruent results have been reported by studies where the metric was mainly used for monitoring affect-induced temperature changes on facial areas (beyond the nose); for example, Engert *et al.* [8] reported a significant drop in chin temperature whereas Veltman and Vos [9] did not. As for the nose, the consistent findings support the fact that mental stressors can cause decreases in its temperature [7]–[9], [15], [16]; however, this thermal directional information itself is still hardly useful in quantifying the amount of stress and as not been tested in unconstrained settings. This suggests that there is a need to explore and build a richer set of metrics to compensate for the limited capability of a single metric, which is likely to lose important physiological information of one's mental stress. This leads to a follow-up research question: *can we build a rich set of metrics to quantify variations in skin temperature?*

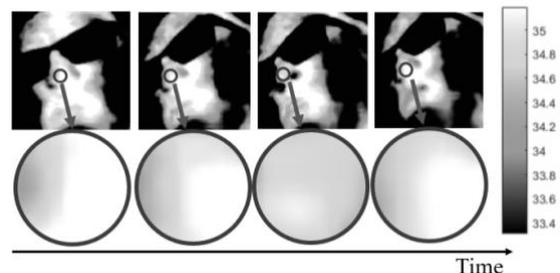

**Fig. 1.** Examples of a person's facial thermal images: the nose tip area on each thermal image has different temperature distribution, not providing common point features across images which are required for the tracking.

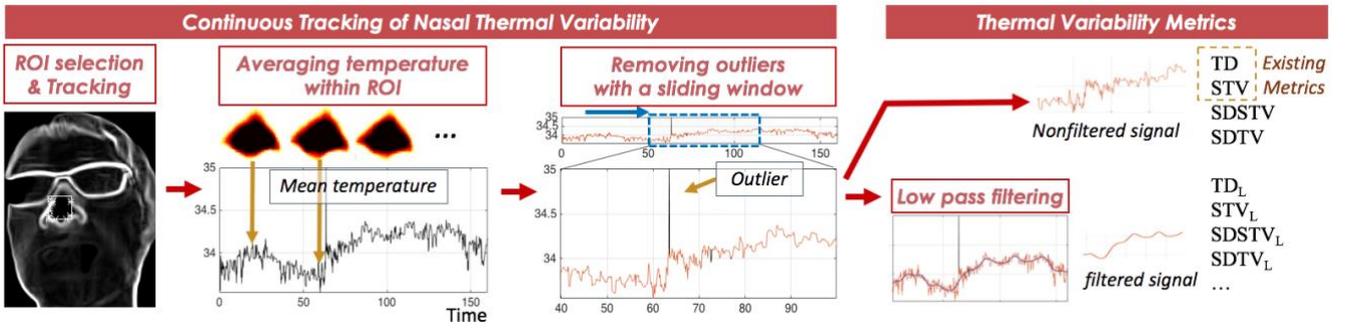

**Fig. 2.** The pipe line of the proposed method for continuous monitoring of dynamic temperature on the nose tip and thermal variability metrics: step a) selecting a larger ROI and tracking, step b) computing average of temperatures within the ROI, step c) removing outliers with a sliding window, d) extracting thermal variability metrics both from low-pass filtered (<0.08Hz) and non-filtered signals.

In the next section, we propose computational methods to address the challenges mentioned above. We first introduce a strategy to track nose temperature continuously. Following this, we propose a set of metrics to capture richer information from the temperature. These new metrics aim to help capture thermal variability rather than only the thermal directionality on which the existing metric is based [7]–[10], [15], [16].

## II. PROPOSED METHOD

Fig. 2 illustrates the pipeline of the proposed method which contains two parts: i) continuous monitoring of nasal variable temperature patterns, ii) extraction of thermal variability metrics.

### A. Continuous Monitoring of Nasal Thermal Variability

Using a thermal camera, an image of a user's face (including the nose) is captured and the nose must be extracted. However, given the blurred and inconsistent shape of the nose tip (skin) on a thermal image in Fig. 1, the conversion of a thermal image to a thermal-gradient map (proposed in [5]) is also less likely to help obtain strong feature points from the area, as shown in Fig. 3-right (there is no facial feature inside the small ROI generally used for tracking the nose tip). Given this, we propose to instead select a larger ROI including the nose tip where a more distinct graphical shape of the nose can be obtained for the tracking as visualized in Fig.3-right. The selection of a larger ROI has the advantage that it can be tracked by advanced thermal ROI tracking algorithms. In this paper, we used Thermal Gradient Flow [5] (given its high performance) to continuously track the area and estimate the nose tip temperature. Accordingly, our first study aims to investigate if the use of a larger ROI size (compared to small ROI size) affects the temperature measurements of the nose tip.

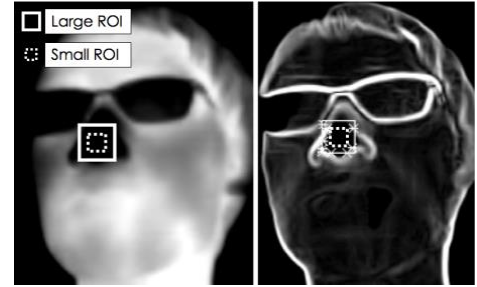

**Fig. 3.** Selection of a larger ROI including the nose tip (small ROI) to obtain strong facial features to enable reliable ROI tracking: (**left**) thermal image, (**right**) thermal gradient map. White stars (*) represent facial point features.

Following [7], [8], [16], we compute the spatial average of temperatures in the ROI from every single frame to obtain a one-dimensional time series of thermal data (Fig. 2). Even if a ROI tracking method achieves high performance in tracking, temporary errors (i.e. tracking failure for a couple of frames) are likely to occur due to the presence of blurred images to which the low-resolution, low-cost thermal imaging are prone to producing (e.g. lens-inducing errors, calibration errors produced by a thermal camera [5]). Hence, the next step is to remove such outliers related to temporary tracking errors as shown in Fig.2-middle. This can be done by excluding values beyond a range computed from Tukey's hinge ($g$=1.5) [18], which has been widely used in outlier rejection. The range can be computed as

$$[Q_1 - g(Q_3 - Q_1), Q_3 + g(Q_3 - Q_1)] \quad (1)$$

where $Q_1$ and $Q_3$ are the first and the third quartiles from temperature distribution and $g$ is the Tukey's constant. To compute the range and remove the outliers, we use a sliding window. In the following studies, we set the length of this window to one third of the total length of each thermal data with the minimum length of 30s.

### B. Thermal Variability Metrics

Studies have shown that nose tip temperature can be affected by not only vasoconstriction-related cardiovascular responses but also respiratory activities [6], [9]. Based on this, we take two types of thermal variable signals from the tracked signal as illustrated in Fig. 2-right: nonfiltered signal (affected by both activities) and low-pass filtered signal for taking relatively slow vasoconstriction/dilation responses (cut-off frequency of 0.08Hz lower than the lower boundary of the expected respiratory range in healthy people [6]).

From both signals, we extract a possible set of metrics that can help capture stress-induced nasal thermal variability including directional information. Here, we present four basic forms of metrics to represent both thermal variability and directionality. The first metric is derived from the existing, widely used metric, *temperature difference* [7], [10], [11], [15], [16]:

| TD | **T**emperature **D**ifference between data from the start and the end |
|---|---|
| | Suppose a thermal variable signal $x(k)$, $k \in [0, n-1]$, $$x(n-1) - x(0)$$ |

The second metric is a slope to capture a global thermal directional trend calculated by a linear polynomial fitting (used in [8], [9]):

| STV | **S**lope of **T**hermal **V**ariable signal |
|---|---|
| | $\beta_1 = \dfrac{y - \beta_0 - \varepsilon}{x}$  (from linear polynomial fitting) |

Inspired by HRV (Heart Rate Variability) metrics [19], two further basic metrics are used to capture physiological thermal variability:

| SDSTV | **S**tandard **D**eviation of **S**uccessive differences of **T**hermal **V**ariable signal |
|---|---|
| | Suppose $\hat{x}(k) = x(k+1) - x(k)$, $$\sqrt{\dfrac{\sum (\hat{x} - \bar{\hat{x}})^2}{n-1}}$$ |

| SDTV | **S**tandard **D**eviation of **T**hermal **V**ariable signal |
|---|---|
| | $$\sqrt{\dfrac{\sum (x - \bar{x})^2}{n-1}}$$ |

We apply the four metrics to both nonfiltered and low-pass filtered signals. As in [6], [20], normalization (min-max feature scaling-based) for each person is considered in this work to further use signals where interpersonal physiological differences are minimized. This leads to a set of sixteen metrics as summarized in Table I, in turn helping take into consideration different perspectives of thermal variability beyond thermal directionality still being the one mainly explored in the literature. The proposed set of metrics are investigated in our second study where psychosocial stress-induction tasks are conducted. Finally, these metrics and the suitability of the media are analyzed in a small third study on a manufacturing shopfloor in a real-life work setting.

## III. DATA COLLECTION STUDIES

The main aim of the data collection studies is two-fold: i) to systematically evaluate the use of a large ROI (versus small) to measure the nose temperatures on thermal videos collected in carefully controlled situations and ii) to verify the capability of the proposed metrics in quantifying mental stress in less constrained settings using stress induction sedentary tasks. Studies were approved by the local University Ethics committee. Prior to data acquisition for each study, participants were given the information sheet and informed consent form.

TABLE I. A NOVEL SET OF METRICS TO QUANTIFY THERMAL VARIABILITY

| Source type | | Nonfiltered | Low-pass Filtered |
|---|---|---|---|
| Original | | **TD**<br>**STV**<br>SDSTV<br>SDTV | $TD_L$<br>$STV_L$<br>$SDSTV_L$<br>$SDTV_L$ |
| Normalized | | $TD_n$<br>$STV_n$<br>$SDSTV_n$<br>$SDTV_n$ | $TD_{Ln}$<br>$STV_{Ln}$<br>$SDSTV_{Ln}$<br>$SDTV_{Ln}$ |

**Bold**: Existing metrics (dominantly used in the literature)

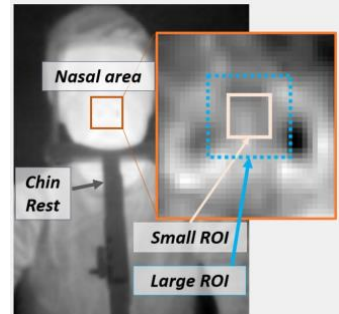

**Fig. 4.** Experimental setup (the image was taken by a thermal camera) for the data collection for Study I: a chin rest was used to maintain the position of the nose of each participant still. The small ROI (reference) includes only the nose tip and the large ROI contains the nose tip and its surrounding area (e.g. part of the nostrils).

### A. Study I: ROI Coverage and Nasal Temperature (N=10)

Fig. 4 shows the experimental setup for the data collection (the image was taken by a thermal camera). The aim was to systematically compare both nonfiltered and filtered thermal variable signals produced from the selection of a large ROI with reference signals from a small ROI containing the nose tip only. We also aimed to investigate the effect of breathing on measurements of nasal temperature through the both ROIs. Hence, participants' mobility had to be carefully controlled to minimize effects of motion-induced noises on nasal temperature measurements even though the automated tracker was still used in the proposed method (the pipeline in Fig. 2). Following [9], we used a chin rest on which each participant relaxed in order to maintain the position of her/his nose to remain as still as possible.

10 healthy adults (aged 22-50years, 4 females) participated in this study that took place in a quiet lab room with no distractions (and no room temperature control). Thermal image sequences were recorded using a low-cost thermal camera (FLIR One 2G) integrated into a smartphone which was placed in front of each participant (circa 50cm).

### B. Study II: Stress Induction Task using Mathematical Serial Subtraction (N=12)

In this study, we used the mathematical serial subtraction task (denoted as *Math*) [21] to induce mental stress for the stress induction. We invited 12 healthy adults (6 females) (aged 18-60 years) from the subject pool service of University College London. Ambient temperature and participants' movements were not controlled.

This task was divided into a resting period, experimental (*Math Hard*) and control (*Math Easy*) sessions as illustrated in Fig. 5. Before starting the experimental and control sessions, participants were asked to rest for 5 minutes. The experimental condition required the participants to repeatedly subtract a two-digit number from a four-digit number for 5 minutes. As introduced in [6], three external stressors were used to ensure induction of psychosocial stress – they are *social evaluative threats*, *time pressure*, *unpleasant sound feedback for wrong answers*. These stressors were introduced because often used in the literature to ensure stress, but also because the literature only compare control vs a stressing condition without exploring the changes in temperature between two different levels of stress. In the control session, participants were asked to countdown mentally with the aim of inducing significantly lower stress levels. Both control and experimental sessions were counterbalanced. Between the sessions, participants were asked to take a break to fully recover from previous session effects. Every participant self-reported his/her perceived stress levels on a visual analogue scale (VAS, 10cm) after each session. The thermal camera in Study I was used to collect participants' facial thermal videos.

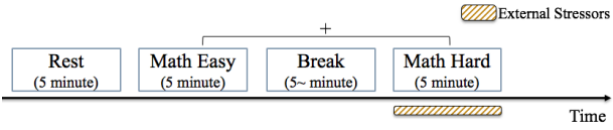

**Fig. 5.** Task flow for Study II: stress induction task consisting of *Rest*, *Math Easy* (controlled) and *Math Hard* (experimental). During the *Math Hard* session, further external stressors were used: social evaluative threats, time pressure, unpleasant sound feedback for wrong answers. The symbol + means *counterbalanced*.

## IV. RESULTS

### A. Comparison between Small and Large ROIs

From Study I with 10 participants, we collected four sets (nonfiltered and filtered signals from both small and large ROIs) of thermal variable signals of 100s. By resampling, we extracted 4000 samples (4 sets x 10 participants x 100 temperature samples from the resampled signals of 100s). The size of each chosen ROI was: i) height (M=8.6 pixels, SD=2.17), width (M=8.5 pixels, SD=2.37) for Small ROI (only on the nose tip); ii) height (16.2 pixels, SD=4.39), width (M=23.7 pixels, SD=7.82) for Large ROI (including the nose tip and its surrounding area). Fig. 6 shows an example of the four sets taken from one participant's thermal images (P2), which show decreasing nasal temperature (affected by colder room temperature). Here, we correlated each nonfiltered thermal samples from a small and a large ROI and then tested correlations between each filtered sample from both ROIs. Overall, both data from the large ROI maintained a high correlation with the data from the small ROI (r=0.999, p<0.001 for both cases).

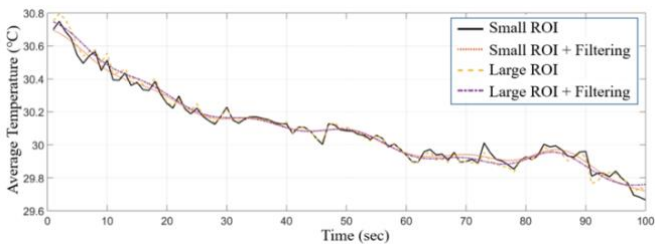

**Fig. 6.** An example of extracted nonfiltered and filtered thermal variable signals from small and large ROIs on facial thermal images of Participant 2.

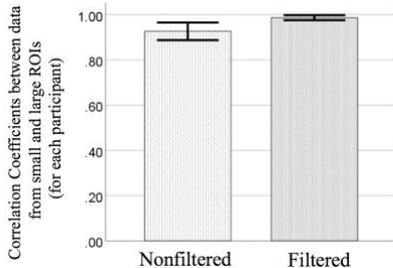

**Fig. 7.** Bar plots of Pearson correlation coefficients between each pair of samples (both data from small and large ROIs) for every individual (10 participants). Each coefficient was computed from each participant's samples (N=100)): (a) nonfiltered data, (b) filtered data.

As a wide range of temperatures from every participant could bring power in achieving high correlation coefficients, we took a look at individual data and correlated them from each pair (small and large ROIs) for every participant. From this, we obtained 10 Pearson correlation coefficients from all 10 participants (we correlated each pair of 100 samples for each participant). As shown in Fig. 7, the results from correlations of individual data still show high levels of correlations for both the filtered and nonfiltered signals from the small and large ROIs. Filtered signals, which should be less affected by breathing artefacts, show indeed stronger correlations between both ROIs (Filtered: M=0.987, SD=0.016; Nonfiltered: M=0.926, SD=0.055).

Furthermore, we explored levels of respiratory effects on nonfiltered nasal thermal signal using the relative power SQI (pSQI) [6]:

$$P(\hat{f}_{min} \leq f \leq \hat{f}_{max}) \cong \frac{\int_{\hat{f}_{min}}^{\hat{f}_{max}} S_x(f)df}{\int_{total} S_x(f)df} \quad (2)$$

where $S_x$ is the power spectral density of signal $x(k)$ and $\hat{f}_{min}$, $\hat{f}_{max}$ are the lower and upper boundary of expected breathing rate (set to [0.1Hz, 0.85Hz] as in [6]) and $0 \leq P \leq 1$. Nonfiltered signals from both ROIs were strongly affected by respiratory cyclic events (respiratory pSQI from the small ROI: M=0.681, SD=0.058; from the large ROI: M=0.692, SD=0.127). This indicates that nasal temperature is affected by breathing regardless of the ROI size.

### B. Effect of Stressors on Thermal Variability Metrics

For Study II, where we mainly aimed to investigate our proposed metrics during stress-induction tasks, we first analyzed self-reported perceived stress scores to test if the harder arithmetic task with external stressors induced significantly higher stress levels than the controlled session and resting period. Boxplots in Fig. 8 show the distributions of the self-reported scores across each session. Overall, it is clear that the stress elicitation procedure produced significantly higher levels of self-reported stress under the experimental condition (*Math Hard* with external stressors) than the ones from the other conditions (*Rest*: M=1.46, SD=1.99; *Math Easy*: M=2.63, SD=1.68; *Math Hard*: M=5.43, SD=2.74), despite having one outlier from one participant (P7, Rest) who felt stressed when rested. We carried out a one-way repeated measures ANOVA on the 12 participants' scores. We found significant differences between each session (F(2,22)=14.792, p<0.001). A post-hoc paired t-test with the Bonferroni correction showed the experimental session (*Math Hard* with external stressors) elicited significantly higher stress levels than other tasks (*Rest – Math Hard*: p=0.005; *Math Easy – Math Hard*: p=0.001). On the other hand, there was no significant effect of the session type on participants' VAS scores over the resting and easy math sessions (p=0.388), indicating other components involved in the tasks (e.g. using a mouse to answer questions) did not significantly affect stress levels.

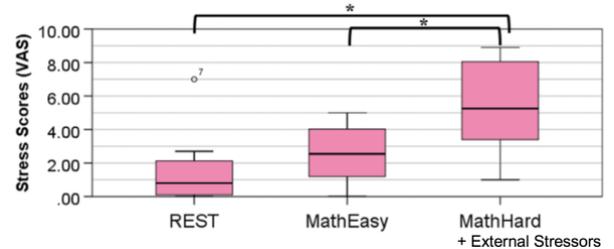

**Fig. 8.** Boxplots (95% confidence interval) of self-reported perceived mental stress scores of 12 participants across each task (Rest, Math Easy, Math Hard with external stressors). The symbol * describes significant differences (p<.05). o[7] (P7, Rest) is a statistical outlier.

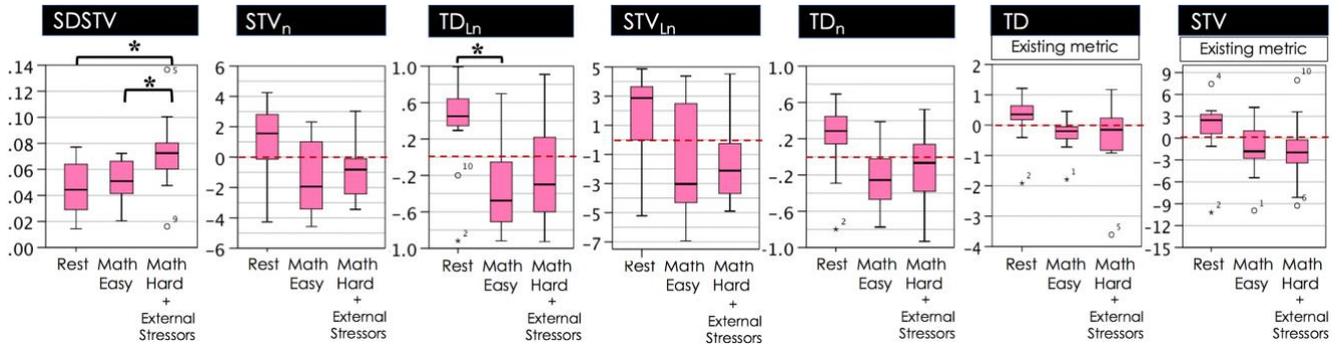

**Fig. 9.** Boxplots (95% confidence interval) of metrics values of 12 participants across tasks (Rest, Math Easy, Math Hard). The symbol * indicates significant differences (p<.05). SDSTV: Standard Deviation of Successive differences of Thermal Variable signal, STV: Slope of Thermal Variable signal, $STV_n$: normalized STV, $STV_{Ln}$: low-pass filtered, normalized STV, TD: Temperature Difference, $TD_{Ln}$: low-pass filtered, normalized TD, $TD_n$: normalized TD. Red dotted lines were added to highlight positive and negative trends for metrics capturing thermal directionality.

For testing the effect of a session type on the proposed set of metrics in Table I, we gathered 576 metrics values (12 participants x 3 sessions x 16 metrics) from facial thermal videos of 12 participants. The ROI tracking for every video was successful (with no tracking failure). We tested the effect of session type on the proposed metrics using a one-way repeated measures ANOVA. Table II summarizes the statistical results and shows a significant effect of the session type on SDSTV from *nonfiltered* signals (F(2,22)=7.053, p=0.004, $\eta_p^2$=0.391), $STV_n$ from *nonfiltered, normalized* signals (F(2,22)=3.619, p=0.044, $\eta_p^2$=0.248) and $TD_{Ln}$ from *low-pass filtered, normalized* signals (F(2,22)=5.575, p=0.011, $\eta_p^2$=0.336). Whilst $TD_n$ from *nonfiltered, normalized* signals and $STV_{Ln}$ from *low-pass filtered, normalized* signals approached significance ($TD_n$: F(2,22)=3.240, p=0.058, $\eta_p^2$=0.228; $STV_{Ln}$: F(2,22)=3.362, p=0.053, $\eta_p^2$=0.234). It should be noticed that the widely used existing metrics, TD and STV from nonfiltered signals showed no significant difference across sessions (TD: p=0.255; STV: p=0.173) although they were generally negative for both *Math Easy* and *Math Hard* (i.e. temperature decreases) and positive for *Rest* (i.e. temperature increases). The results also confirmed normalization helped minimizing interpersonal physiological variances, in turn contributing to statistical significance (e.g. $TD_n$ and TD in Fig. 9).

The post-hoc paired t-test with Bonferroni correction on each metric showed that **SDSTV** for *Math Hard* with external stressors was significantly higher than both *Rest* (p=0.046) and *Math Easy* (p=0.031). SDSTV patterns had a certain level of similarity with self-reported perceived mental stress scores (see Fig. 8 and Fig. 9 SDSTV), indicating the importance of capturing thermal variability in assessing mental stress level. **$TD_{Ln}$** showed a significant difference between *Rest* and *Math Easy* (p=0.04) but with no significant differences between other pairs, indicating that nasal temperature could decline during cognitive activity despite lower levels of perceived mental stress. Interestingly, the amount of decreases in nasal temperature during *Math Hard* was generally lower than *Math Easy* (Fig. 9 $TD_{Ln}$), indicating that metrics for thermal directionality may be less informative in quantifying the amount of mental stress. Other than these two, there were no significant differences for each pair as shown in Fig. 9.

## V. REAL WORLD APPLICATION

*Case Study: Stress Monitoring of Human Workers in Factory Workplace*

A possible application of mobile thermal imaging-based stress monitoring strategy is to use it as an assistive technology to improve the capability and wellbeing of the human workforce by tailoring work schedules or activities towards her/his psychological needs [22]. These needs are likely to be reflected by the amount of mental loading and stress that a worker experiences. For example, when the system detects the factory worker feeling very stressed whilst assembling, it can provide more detailed computer-aided assembly instructions (e.g. on a virtual reality headset).

To assess the capability and feasibility of the proposed method beyond laboratory situations, we visited a factory workplace (ROYO, Spain - furniture manufacturer) and had three skilled workers to use our monitoring system in their routinely furniture assembly tasks. To place a thermal camera

TABLE II. SIGNIFICANCE TEST TO ASSESS THE EFFECT OF A SESSION TYPE ON EACH METRIC USING A ONE-WAY REPEATED MEASURES ANOVA

| Source | Measure | df | F | p-value | Partial Eta squared |
|---|---|---|---|---|---|
| Task | $TD^e$ | 2 | 1.456 | .255 | .117 |
| | $STV^e$ | 2 | 1.900 | .173 | .147 |
| | **SDSTV*** | 2 | 7.053 | **.004** | **.391** |
| | SDTV | 2 | 1.194 | .322 | .098 |
| | $TD_n^\pm$ | 2 | 3.240 | .058 | .228 |
| | **$STV_n$*** | 2 | 3.619 | **.044** | **.248** |
| | $SDSTV_n$ | 2 | 1.023 | .376 | .085 |
| | $SDTV_n$ | 2 | 1.236 | .310 | .101 |
| | $TD_L$ | 2 | 2.235 | .131 | .169 |
| | $STV_L$ | 2 | 1.908 | .172 | .148 |
| | $SDSTV_L$ | 2 | 1.188 | .324 | .097 |
| | $SDTV_L$ | 2 | .469 | .632 | .041 |
| | **$TD_{Ln}$*** | 2 | 5.575 | **.011** | **.336** |
| | $STV_{Ln}^\pm$ | 2 | 3.362 | .053 | .234 |
| | $SDSTV_{Ln}$ | 2 | .140 | .870 | .013 |
| | $SDTV_{Ln}$ | 2 | 2.217 | .133 | .168 |

* significant, ± approaching significant, e existing metrics

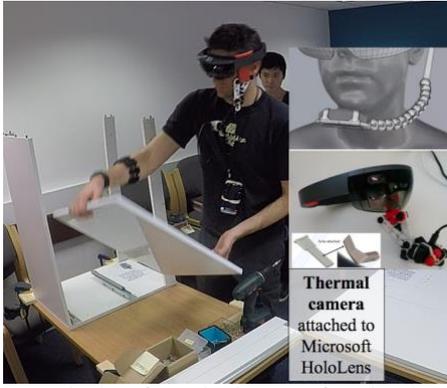

**Fig. 10.** Stress monitoring of a worker in a manufacturing shopfloor, as part of EU H2020 Human project. A low-cost thermal camera was attached to an MR headset (Microsoft HoloLens).

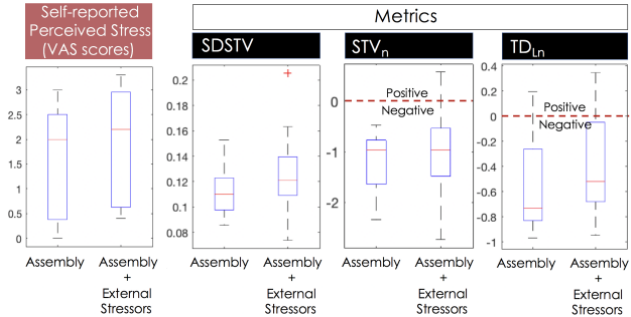

**Fig. 11.** Boxplots (95% confidence interval) of three metrics (SDSTV, $STV_n$, $TD_{Ln}$) values from three factory workers across two type of tasks (Assembly, Assembly with external stressors. In total 24 instances).

near the face during the physical activity, we built a headset-shaped interface shown in Fig. 10, which can also be used together with wearable devices, e.g. Microsoft mixed reality (MR) headset HoloLens in our case. With this setup, we collected 12 thermal videos from the workers during their normal tasks. Given written consents from the manufacturer and workers, we presented external stressors (social evaluative threats by performance observers, time pressure proposed in [6]) during their assembly activities, resulting in producing further 12 thermal videos of the workers with higher stress levels self-reported on 10cm-VAS (Fig. 11). We applied our proposed method to the collected thermal recordings to compute the three metrics that showed significant differences across tasks in the previous section (SDSTV, $STV_n$, $TD_{Ln}$). As shown in Fig. 11, SDSTV values for the assembly task with external stressors (mean: 0.13, SD: 0.03) were generally higher than data collected during the normal assembly task (mean: 0.11, SD: 0.02), which is similar to the results in Study II (see Fig. 9 left). For $STV_n$, $TD_{Ln}$, their values were generally negative for both sessions (nasal temperature decreased). Interestingly, as in Study II, decreases in nasal temperature were less strong for the session with external stressors than the other although stastical analysis cannot be carried out on this small set of participants.

## VI. DISCUSSION AND CONCLUSION

This paper has aimed to contribute to the body of work in thermal imaging-based affective computing [6]–[10], [23]–[26] by building methods that can continuously and reliably capture richer information of stress-induced nasal thermal variability through mobile thermal imaging.

Despite the importance of using a large ROI (the nose tip with its surrounding areas) as a surrogate of the only nose tip for automatic motion tracking in real world applications, the effect of different ROI coverages on nasal temperature measurements has not been explored in the literature [7]–[9], [15], [16]. From Study I, we have found strong correlations between data derived from the large and small (nose tip only) ROIs. This is highly encouraging in that it could be possible to deploy this approach to real world contexts through more robust tracking techniques. We have also confirmed that respiration influences nasal temperature regardless of the ROI size; hence, there is a need to take into account respiratory and vasoconstriction/dilation activities separately when monitoring nasal temperature.

Capturing multiple aspects of physiological variability has been shown to be important in assessing mental stress levels; for example, multiple metrics have been proposed to capture heart rate variability related to mental states [19]. However, this has not yet been explored for physiological thermal variability. Therefore, we have proposed a novel set of metrics with the aim to capture complex aspects of the physiological phenomenon beyond just the dominantly used thermal directionality. Also, our studies investigated this in the experimental settings comparing different levels of stress (control and experimental) against the baseline. Our results showed that variability-based metrics were more helpful in quantifying stress levels than the typically used directionality one. Indeed, whilst nasal temperatures generally declined when a person carried on both sedentary cognitive tasks, thermal directional cues from the two existing metrics (TD, STV in Fig. 9) were not sufficiently sensitive to mental stress levels. This was also the case for normalization (to consider physiological interpersonal variability) and normalization with low-pass filtering although we observed increases in the effect sizes (TD: $\eta_p^2=.117$, $TD_n$: $\eta_p^2=.228$; $TD_{Ln}$: $\eta_p^2=.336$). On the other hand, one of the metrics to capture variability of both vasoconstriction/dilation and respiration, SDSTV, was highly informative of mental stress with high levels of similarity with perceived mental stress scores (Fig. 8 and 9).

Furthermore, we also observed that whilst, nasal temperatures during the more stressful sessions (Math Hard + social pressure; Assembly + social pressure) were less decreased than the less stressful ones (Math Easy; Assembly only). This could be explained by the interaction of the different types of stress (mental overload vs. social pressure). Social pressure (in our case, imposing social evaluative threats) may cause embarrassment that has been shown to lead to increase (rather than decrease) in temperature [17]. Although such thermal responses to embarrassment have been mainly studies in interpersonal touch, still these results call for developing thermal metrics that capture patterns rather than direction to take into account both the physiological phenomena but also the interaction of multiple factors if they have to be used in everyday life. All in all, the proposed set of metrics could compensate for this, highlighting the importance of capturing richer information of thermal variability so as to avoid collapsing its complex phenomena into a single metric.

Despite the above findings, there are still limitations opening up future research opportunities. First, we have used the spatial averaging method to summarize two dimensional thermal information on the nasal area. This could lead to losing important vasoconstriction/dilation induced local thermal variances [5]. Second, despite our proposed metrics helping quantify mental stress from multiple perspectives, even carefully hand-engineered metrics can hardly capture all complex aspects of signals, particularly in in-the-wild situations [6]. By addressing them, we expect that the *Nose Heat* approach can support a wider range of real-world applications as an affective assistive technology.

ACKNOWLEDGMENT

We wish to acknowledge that the real-world case study was supported by the EU Horizon 2020 HuMan project (http://www.humanmanufacturing.eu/)

REFERENCES

[1] E. F. J. Ring and K. Ammer, 'Infrared thermal imaging in medicine', *Physiol. Meas.*, vol. 33, no. 3, p. R33, 2012.
[2] E. F. J. Ring, 'The historical development of thermal imaging in medicine', *Rheumatology (Oxford)*, vol. 43, no. 6, pp. 800–802, Jun. 2004.
[3] L. J. Jiang *et al.*, 'A perspective on medical infrared imaging', *Journal of Medical Engineering & Technology*, vol. 29, no. 6, pp. 257–267, Jan. 2005.
[4] Y. Cho, N. Bianchi-Berthouze, N. Marquardt, and S. J. Julier, 'Deep Thermal Imaging: Proximate Material Type Recognition in the Wild Through Deep Learning of Spatial Surface Temperature Patterns', in *Proceedings of the 2018 CHI Conference on Human Factors in Computing Systems*, New York, NY, USA, 2018, pp. 2:1–2:13.
[5] Y. Cho, S. J. Julier, N. Marquardt, and N. Bianchi-Berthouze, 'Robust tracking of respiratory rate in high-dynamic range scenes using mobile thermal imaging', *Biomed. Opt. Express, BOE*, vol. 8, no. 10, pp. 4480–4503, Oct. 2017.
[6] Y. Cho, S. J. Julier, and N. Bianchi-Berthouze, 'Instant Stress: Detection of Perceived Mental Stress Through Smartphone Photoplethysmography and Thermal Imaging', *JMIR Mental Health*, vol. 6, no. 4, p. e10140, 2019.
[7] C. K. L. Or and V. G. Duffy, 'Development of a facial skin temperature-based methodology for non-intrusive mental workload measurement', *Occupational Ergonomics*, vol. 7, no. 2, pp. 83–94, Jan. 2007.
[8] V. Engert, A. Merla, J. A. Grant, D. Cardone, A. Tusche, and T. Singer, 'Exploring the Use of Thermal Infrared Imaging in Human Stress Research', *PLOS ONE*, vol. 9, no. 3, p. e90782, Mar. 2014.
[9] J. A. Veltman and W. K. Vos, 'Facial temperature as a measure of mental workload', in *International Symposium on Aviation Psychology*, Oklahoma City, 2005, vol. pp.777-781.
[10] A. Di Giacinto, M. Brunetti, G. Sepede, A. Ferretti, and A. Merla, 'Thermal signature of fear conditioning in mild post traumatic stress disorder', *Neuroscience*, vol. 266, pp. 216–223, Apr. 2014.
[11] E. Salazar-López *et al.*, 'The mental and subjective skin: Emotion, empathy, feelings and thermography', *Consciousness and Cognition*, vol. 34, pp. 149–162, Jul. 2015.
[12] Y. Cho, N. Bianchi-Berthouze, and S. J. Julier, 'DeepBreath: Deep Learning of Breathing Patterns for Automatic Stress Recognition using Low-Cost Thermal Imaging in Unconstrained Settings', in *the 7th International Conference on Affective Computing and Intelligent Interaction, ACII 2017*, 2017, pp. 456–463.
[13] P. E. Pergola, D. L. Kellogg, J. M. Johnson, W. A. Kosiba, and D. E. Solomon, 'Role of sympathetic nerves in the vascular effects of local temperature in human forearm skin', *American Journal of Physiology-Heart and Circulatory Physiology*, vol. 265, no. 3, pp. H785–H792, Sep. 1993.
[14] M. Elam and B. G. Wallin, 'Skin blood flow responses to mental stress in man depend on body temperature', *Acta Physiologica Scandinavica*, vol. 129, no. 3, pp. 429–431, Mar. 1987.
[15] H. Genno *et al.*, 'Using facial skin temperature to objectively evaluate sensations', *International Journal of Industrial Ergonomics*, vol. 19, no. 2, pp. 161–171, Feb. 1997.
[16] Y. Abdelrahman, E. Velloso, T. Dingler, A. Schmidt, and F. Vetere, 'Cognitive heat: exploring the usage of thermal imaging to unobtrusively estimate cognitive load', *Proceedings of the ACM on Interactive, Mobile, Wearable and Ubiquitous Technologies*, vol. 1, no. 3, p. 33, 2017.
[17] A. C. Hahn, R. D. Whitehead, M. Albrecht, C. E. Lefevre, and D. I. Perrett, 'Hot or not? Thermal reactions to social contact', *Biology letters*, vol. 8, no. 5, pp. 864–867, 2012.
[18] J. Tukey, *Exploratory Data Analysis*. Pearson, 1977.
[19] F. Shaffer and J. P. Ginsberg, 'An Overview of Heart Rate Variability Metrics and Norms', *Front Public Health*, vol. 5, Sep. 2017.
[20] J. Hernandez, R. R. Morris, and R. W. Picard, 'Call Center Stress Recognition with Person-Specific Models', in *Affective Computing and Intelligent Interaction*, 2011, pp. 125–134.
[21] R. Soufer *et al.*, 'Cerebral cortical hyperactivation in response to mental stress in patients with coronary artery disease', *PNAS*, vol. 95, no. 11, pp. 6454–6459, May 1998.
[22] Y. Cho, 'Automated Mental Stress Recognition through Mobile Thermal Imaging', in *the 7th International Conference on Affective Computing and Intelligent Interaction, ACII 2017*, 2017, pp. 596–600.
[23] I. Pavlidis, N. L. Eberhardt, and J. A. Levine, 'Human behaviour: Seeing through the face of deception', *Nature*, vol. 415, no. 6867, pp. 35–35, Jan. 2002.
[24] L. Gane, S. Power, A. Kushki, and T. Chau, 'Thermal Imaging of the Periorbital Regions during the Presentation of an Auditory Startle Stimulus', *PLOS ONE*, vol. 6, no. 11, p. e27268, Nov. 2011.
[25] D. A. Pollina *et al.*, 'Facial Skin Surface Temperature Changes During a "Concealed Information" Test', *Ann Biomed Eng*, vol. 34, no. 7, pp. 1182–1189, Jun. 2006.
[26] S. J. Ebisch, T. Aureli, D. Bafunno, D. Cardone, G. L. Romani, and A. Merla, 'Mother and child in synchrony: Thermal facial imprints of autonomic contagion', *Biological Psychology*, vol. 89, no. 1, pp. 123–129, Jan. 2012.